\documentstyle{article}
\newcommand{\bm}[1]{\mbox{\boldmath $#1$}}
\newcommand{\be}{\begin{equation}}
\newcommand{\ee}{\end{equation}}
\newcommand{\bea}{\begin{eqnarray}}
\newcommand{\eea}{\end{eqnarray}}
\newcommand{\st}{{\scriptscriptstyle T}}
\newcommand{\xbj}{x_{\scriptscriptstyle B}}

\newcommand{\tv}[1]{{\bm #1}_{\st}}
\newcommand{\bSt}{\tv{S}}

\begin{document}
\input{epsf}
\title{
AZIMUTHAL AND SINGLE SPIN ASIMMETRIES IN DEEP INELASTIC SCATTERING  
\protect} 
\vspace{3mm}
\author {M. Boglione and P.J. Mulders\\  
\vspace{3mm}
\mbox{}\\
{\it Division of Physics and Astronomy, Faculty of Science, Free University}\\
{\it De Boelelaan 1081, NL-1081 HV Amsterdam, the Netherlands}\\
}
\date{}
\maketitle

\begin{abstract}
Azimuthal and single spin asymmetries play a crucial role in the study of the 
spin structure of hadrons in terms of their elementary constituents. 
Exploiting them to uncover information over subtle distribution and 
fragmentation functions, which cannot easily be accessed in other ways, is 
what this talk is about.
\end{abstract}

\section{Introduction}
We are familiar with double spin asymmetries in DIS, the most celebrated 
examples being those used to determine the distribution functions $g^a_1$ 
and $g^a_2$:
\be
\frac{
\sigma ^{\stackrel{\rightarrow}{\rightarrow}} - 
\sigma ^{\stackrel{\rightarrow}{\leftarrow}}}
{\sigma ^{\stackrel{\rightarrow}{\rightarrow}} + 
\sigma ^{\stackrel{\rightarrow}{\leftarrow}}}
\sim \,\sum _a e_a^2 g_1^a
\ee
\be
\frac{
\sigma ^{{\rightarrow}{\uparrow}} - 
\sigma ^{{\rightarrow}{\downarrow}}}
{\sigma ^{{\rightarrow}{\uparrow}} + 
\sigma ^{{\rightarrow}{\downarrow}}}
\sim \frac{M}{Q}\,\sum _a e_a^2 (\frac{y}{2}\,g_1^a+g_2^a)\,.
\ee
Single spin asymmetries are similar quantities, where only one 
of the two initial state particles is polarized
\be
\frac
{\sigma ^{\uparrow} - 
\sigma ^{\downarrow}}
{\sigma ^{\uparrow} + 
\sigma ^{\downarrow}} \,
\ee
and give access to other more subtle distribution and fragmentation functions.
Single spin asymmetries are conceptually very simple objects;
nevertheless, results are often striking and could not, until recently, be 
explained in plain pQCD parton model.
Moreover, single spin asymmetries are crucial tools in the study 
of the spin structure of hadrons, since 
they are strictly related to two ``hot'' topics in this field:
\begin{enumerate}
\item The role of the intrinsic transverse momentum $k_T$ of 
the parton relative to that of the parent spin $1/2$ hadron, in both the 
dynamics and 
the kinematics of the process (single spin asymmetries are zero when $k_T$ 
effects are neglected)
\item The existence of T-odd distribution and fragmentation functions, which 
are not forbidden by time reversal invariance 
(single spin asymmetries are zero unless at 
least one of the functions in the correlators is T-odd).
\end{enumerate}
Let's discuss these points in some more detail.

\subsection{The Intrinsic Transverse Momentum}

In most of the cases collinear kinematics is good enough an approximation to 
successfully describe DIS processes and determine distribution functions like 
$f_1^a$, $g_1^a$ and $g_2^a$. In this case the hadron is described as an 
ensamble of quarks moving in the same direction as the parent hadron.

\begin{figure}[h]
\centering
\mbox{\epsfysize=15mm\epsffile{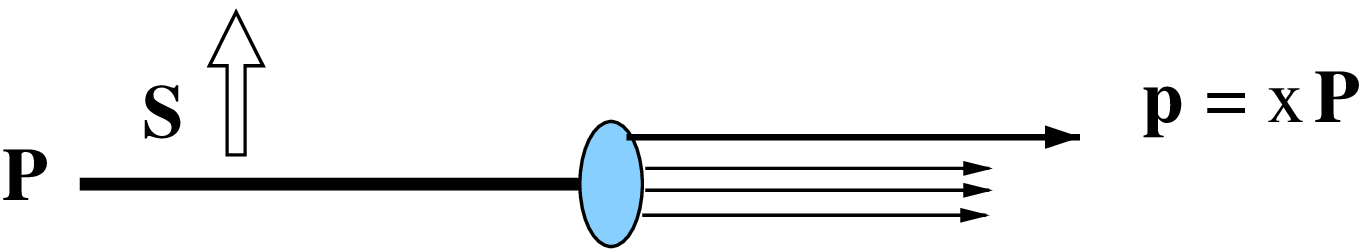}}
\end{figure}

But in other cases, such as azimuthal and single spin asymmetries, 
the approximation of collinear kinematics leads to vanishing results,  
incompatible with experimental data. In these cases it is necessary to 
take into account the contribution of the intrinsic 
transverse momentum of the quarks inside the hadron. See Ref.~\cite{lathuile} 
for more details and plots.

\begin{figure}[h]
\centering
\mbox{\epsfysize=15mm\epsffile{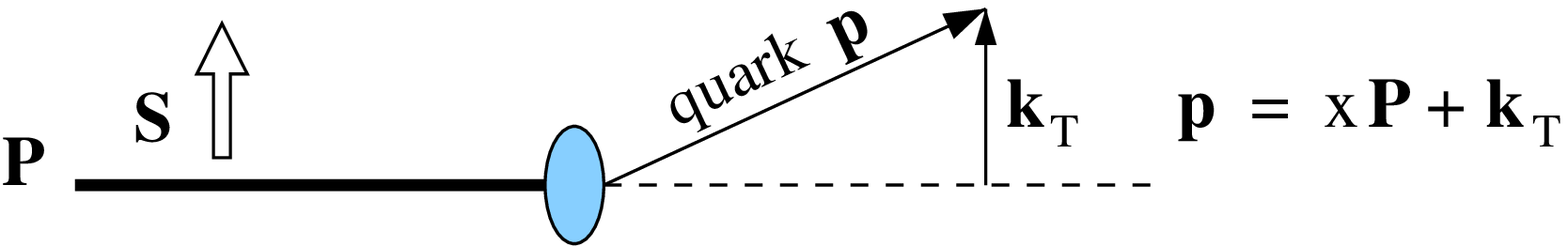}}
\end{figure}

\subsection{Time Reversal Invariance}

Correlation functions fulfill hermiticity, parity and charge conjugation 
invariance. Time-reversal symmetry is a more delicate issue: in this case 
target hadron  and produced hadrons have to be considered separately.
In fact, in the initial state, we have only {\it one} target nucleon
and so only {\it one} momentum characterizes the 
initial hadronic state $|P>_{in}$. Now, if we apply a time reversal 
transformation, we obtain a final state that is exactly the same as the 
initial state, 
\(
|P>_{out} \,=\, |P>_{in}
\).
Therefore the correlator $\Phi$ \cite{Soper77,Jaffe83} which defines the distribution functions is  
time-reversal invariant, at least for DIS processes where initial state 
interactions cannot easily occur.
On the contrary, in the final state we have a number of hadrons produced, and 
so a larger number of momenta characterize the final state vector. Now, if we 
perform a time reversal transformation on this state we obtain a different 
state, which differs from the first one by a phase, 
\(
|P_h,X>_{in} \,=\, e^{i\delta _{hX}}|P_h,X>_{out}
\). 
Therefore the corrrelator $\Delta$, which define the fragmentation functions,  
is not time reversal invariant. As a consequence, T-odd fragmentation 
functions do exist and  play a crucial role:
as we have anticipated, single spin asymmetries are directly related to 
T-odd functions. In fact, they turn out to be zero unless either the 
correlator $\Phi$ or $\Delta$ contains at least one T-odd function.

\section{Single Spin Asymmetry in $p^{\uparrow}p \to \pi X$}

Experimental data on single spin asymmetries are only just starting to be 
available, but some interesting work has already been done \cite{abm}
relying on a very 
accurate measurement of the single spin asymmetry of pions, semi-inclusively 
produced in $p^{\uparrow}p$ scattering \cite{adams}. 
Since
\be
\sigma ^{\uparrow} - \sigma ^{\downarrow} \ \propto 
\ f_1 \otimes h_1 \otimes H_1^\perp\,,
\ee
from a fit of these data we were able to determine \cite{bl00}, through an 
appropriate parameterization which takes into account Soffer 
bounds and positivity constraints, the functions  $h_1$ and $H_1^\perp$. 
The first, $h_1(x)$,  is the well known ``transversity'' distribution 
function, which cannot be measured in DIS due to its chiral odd character 
and has not yet been experimentally accessed in any other process other 
than $p^{\uparrow}p$ scattering (RHIC and HERMES will hopefully have 
new data soon, from which new precious information on $h_1$ can be extracted). 
The second function, $H_1^\perp(z)$, is a chiral odd and $k_T$ dependent 
fragmentation function, which describes the fragmentation of polarized quarks 
into spinless hadrons, such as pions.

\section{Azimuthal and single spin asymmetries in DIS}

The functions $h_1$ and $H_1^\perp$, determined by fitting 
$p^{\uparrow}p \to \pi X$ experimental data, can be used to give interesting 
predictions of DIS azimuthal single spin asymmetries which are presently 
being measured by HERMES and SMC (with longitudinally polarized target) or 
that will be measured in the near future (with transversely polarized target).
See Ref.~\cite{bm98} for definitions and details.
In terms of weighted integrals, for transversely polarized target we have:
\bea
&&\mbox{}\hspace{-1.3cm}
\left<\frac{Q_\st}{M} \,\sin(\phi^\ell_h-\phi^\ell_S)\right>_{OT} 
\nonumber  \\ &&\mbox{}\hspace{-2.0cm}\qquad {}
= \frac{2\pi \alpha^2\,s}{Q^4}\,\vert \bSt \vert
\,\left( 1-y+\frac{1}{2}\,y^2\right)\sum_{a,\bar a} e_a^2
\,\xbj\, f_{1T}^{\perp(1)a}(\xbj) D^a_1(z_h)\,,
\\&&\mbox{}\hspace{-1cm}
\nonumber \\&&\mbox{}\hspace{-1cm}
\left<\frac{Q_\st}{M_h}\,\sin(\phi^\ell_h+\phi^\ell_S)\right>_{OT}
\nonumber \\ &&\mbox{}\hspace{-1cm}\qquad {}
= \frac{2\pi \alpha^2\,s}{Q^4}\,\vert \bSt \vert
\,2(1-y)\sum_{a,\bar a} e_a^2
\,\xbj\,h_1^a(\xbj)H_1^{\perp(1)a}(z_h)\,.
\eea
Predictions for the $x$ and $z$ dependence of these asymmetries are given in 
Ref.~\cite{BM99}
For longitudinally polarized targets we have the following single spin 
asymmetries:
\bea
&&\mbox{}\hspace{-1.0cm}
\left<\frac{Q_\st^2}{4MM_h}\,\sin(2\phi^\ell_h)\right>_{OL}
= -\frac{4\pi \alpha^2\,s}{Q^4}\,\lambda\,(1-y)
\nonumber \\ &&\mbox{}\hspace{0.3cm} \quad \mbox{}
\times \sum_{a,\bar a} e_a^2
\,\xbj\,h_{1L}^{\perp(1)a}(\xbj) H_1^{\perp(1)a}(z_h)\,,
\label{sin2}
\eea
\bea
&&\mbox{}\hspace{-1cm}
\left<\frac{Q_T}{M}\,\sin(\phi^\ell_h)\right>_{OL}
= \frac{4\pi \alpha^2\,s}{Q^4} \,\lambda\,(2-y)\sqrt{1-y}
\,\frac{2M_h}{Q}
\nonumber \\ &&\mbox{}\hspace{-1cm} 
\quad \mbox{}\times\sum_{a,\bar a} e_a^2 \Biggl\{
\xbj h_{1L}^{\perp (1)a}(\xbj)
\,\frac{\tilde H^a(z_h)}{z_h}
\nonumber \\ && \hspace{0.5 cm} \mbox{} +
\xbj \left(2\,h_{1L}^{\perp (1) a} 
- \xbj \, h_L^a (\xbj)
\right) \, H_1^{\perp(1)a}(z_h)
\Biggr\}\,.
\label{sin}
\eea
The fragmentation function $\tilde H$ needed to estimate these weighted 
integrals can be obtained using the relation 
\be
\frac{H^a(z)}{z}=z^2\,\frac{d}{dz}\,
\Bigl( \frac{H_1^{\perp (1) a}(z)}{z} \Bigr),
\ee
which follows from Lorentz covariance.
Unfortunately, it is not possible to give a similar straight forward equation  
to determine the distribution functions $h_{1L}^{\perp (1)}$ and $h_L$ 
appearing in Eq.~\ref{sin2} and \ref{sin}. Nevertheless, they can be estimated 
by using two somehow opposite approximations \cite{BM00}.

The first is to assume that the contribution of the function 
$\tilde h _L (x)$, the interaction dependent term in $h_L$, and the
quark mass terms can be neglected.  This means  
\begin{eqnarray}
h_L (x) &=& 2x \, \int _x^1 dy\, \frac{h_1(y)}{y^2} , \\
h_{1L}^{\perp (1)} (x) &=& -\frac{1}{2} x h_L(x) =
-x^2 \, \int _x^1 dy\, \frac{h_1(y)}{y^2} ,
\end{eqnarray}

The second assumption we consider is that $h_{1L}^{\perp (1)} (x)$ 
is small enough to be neglected (and again quark mass terms are
neglected too). In this approximation we obtain 
\begin{eqnarray}
h_L(x) &=& h_1(x) \,, \\
\tilde h_L(x) &=& h_1(x) \,,\\
\overline h_L(x) &=& h_1(x) - 2x\int _x^1 dy \frac{h_1(y)}{y^2}\,.
\end{eqnarray}

\begin{figure}
\begin{minipage}{6cm}
\mbox{\epsfysize=50mm\epsffile{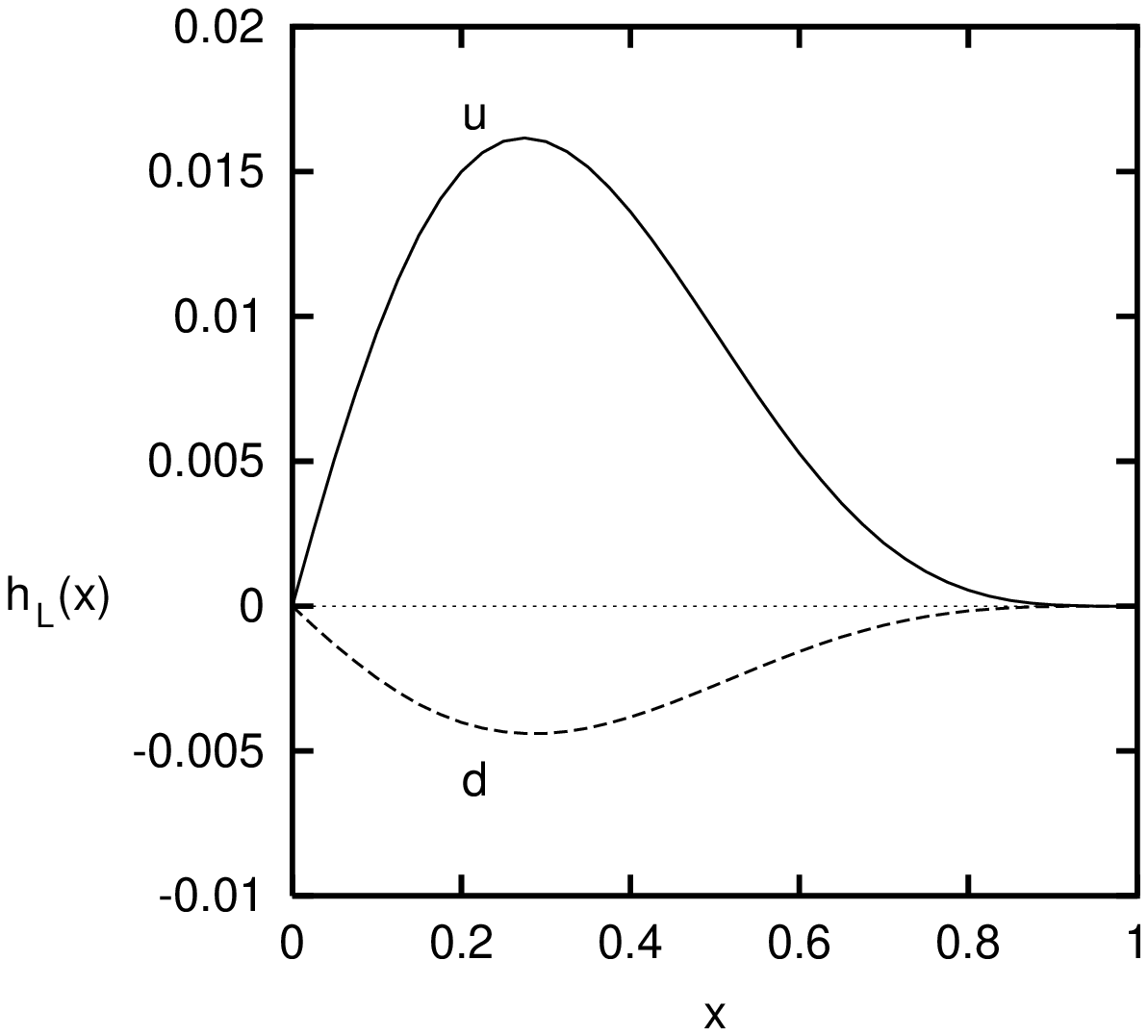}}
\end{minipage}
\begin{minipage}{6cm}
\mbox{\epsfysize=50mm\epsffile{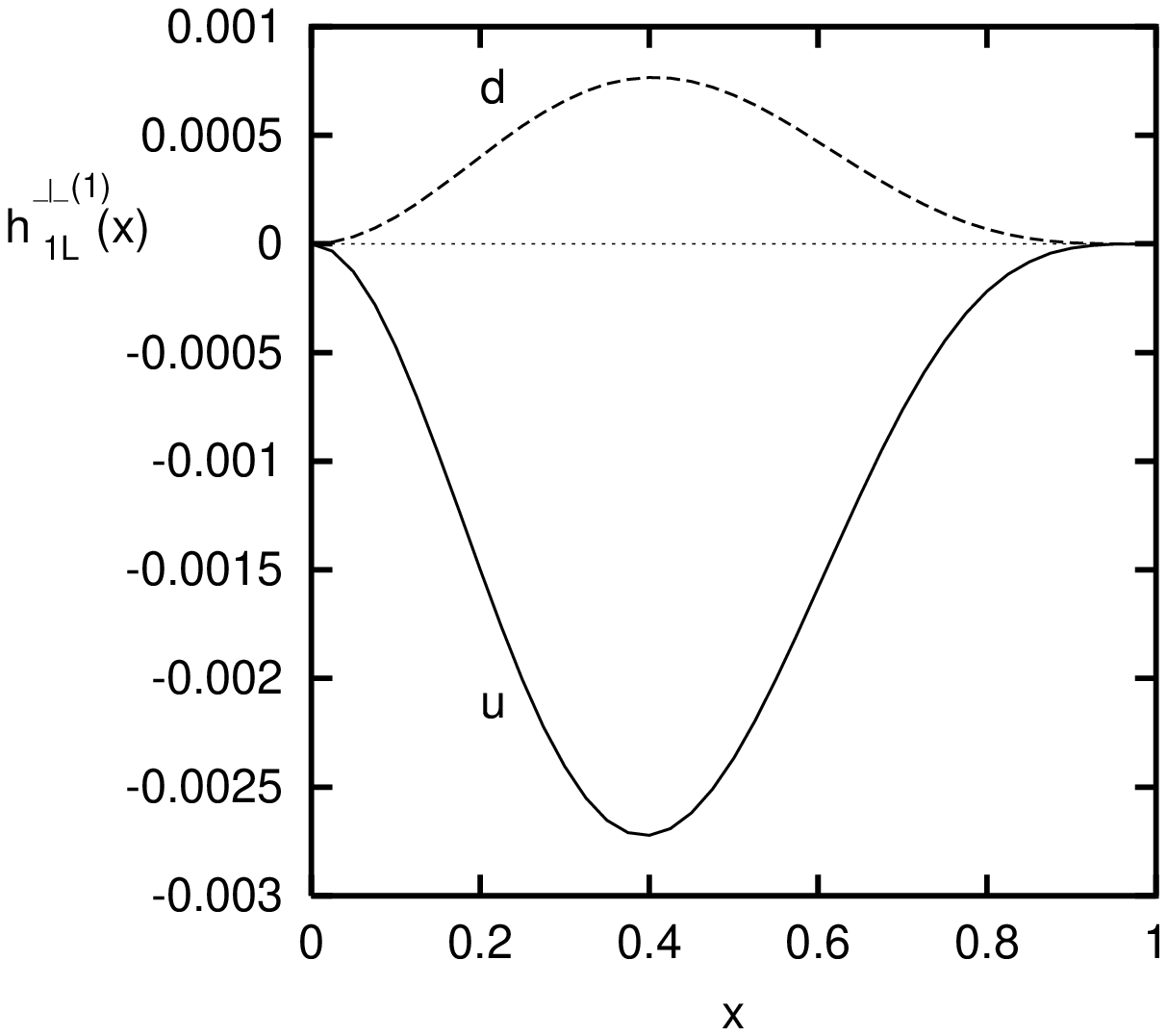}}
\end{minipage}
\caption{ The distribution functions $h_L^u(x)$, $h_L^d(x)$  and 
$h_{1L}^{\perp (1) u}(x)$, $h_{1L}^{\perp (1) d}(x)$, as obtained  
under the approximation $\tilde h_L(x) =0$.}
\label{F1}  
\end{figure}

\begin{figure}
\begin{minipage}{6cm}
\mbox{\epsfysize=50mm\epsffile{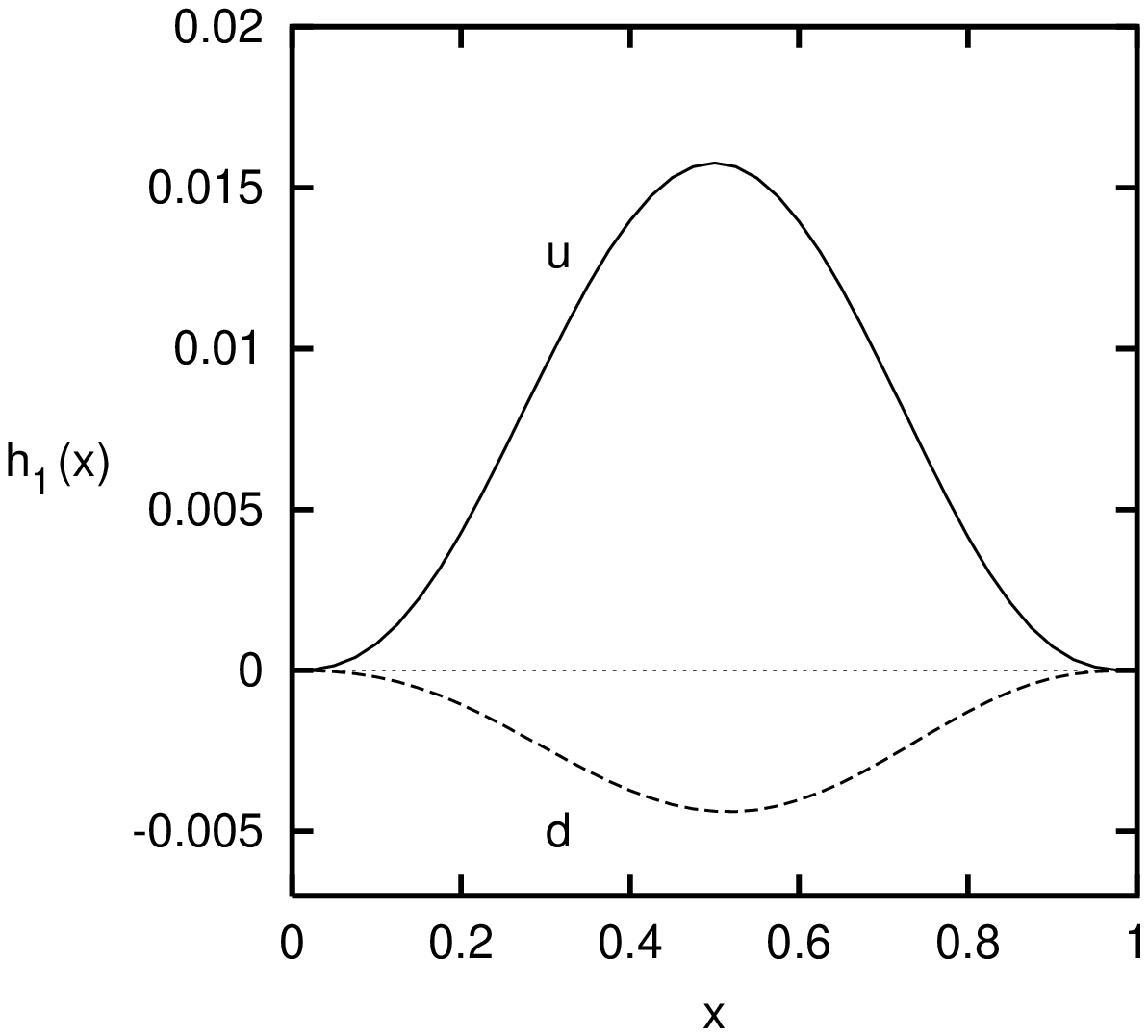}}
\end{minipage}
\begin{minipage}{6cm}
\mbox{\epsfysize=50mm\epsffile{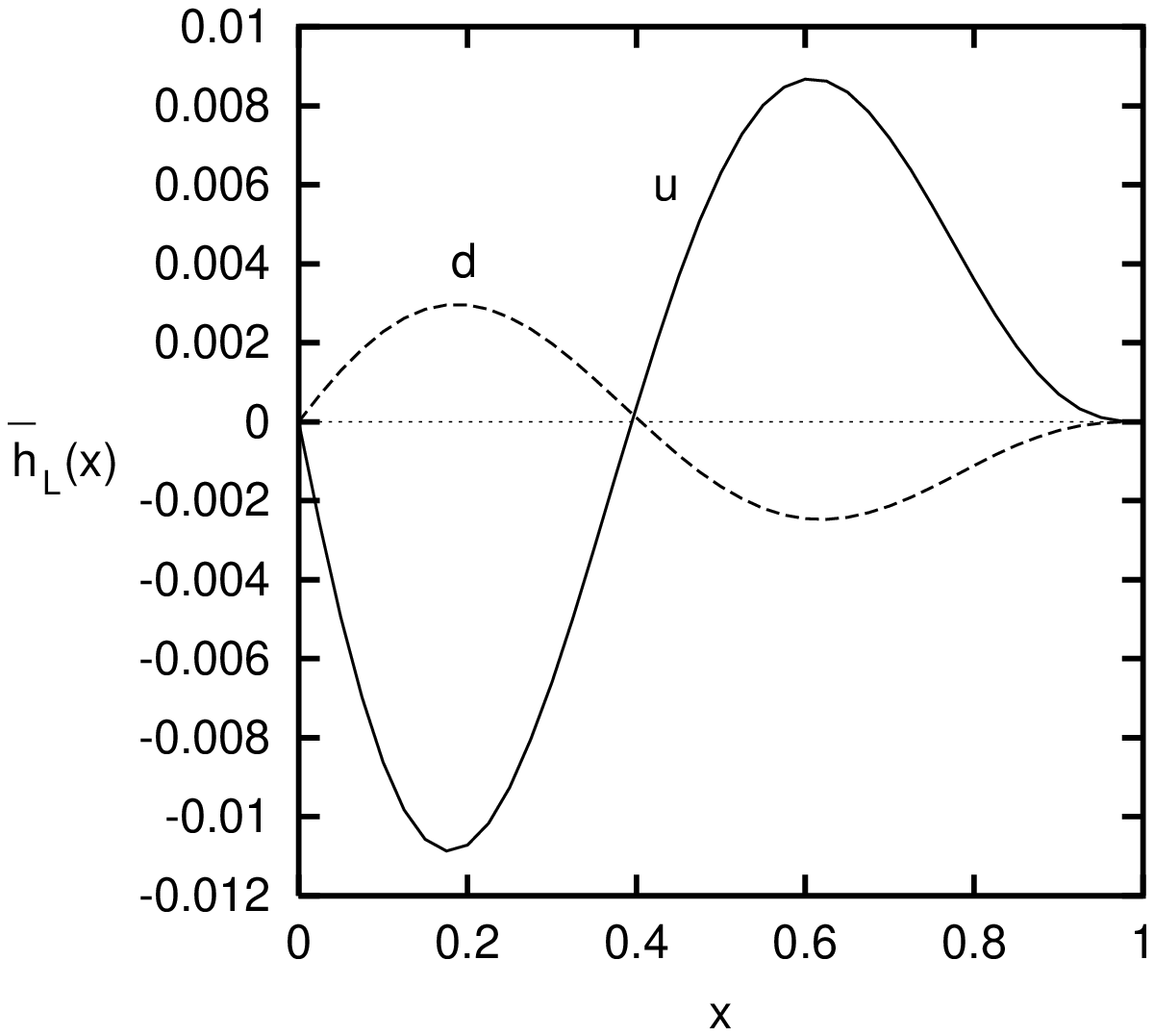}}
\end{minipage}
\caption{ The distribution functions $h_1$ as obtained from the fit on 
$p^{\uparrow} p \to \pi X$ experimental data, and the distribution functions 
$\overline h_L^u(x)$ and $\overline h_L^d(x)$, as obtained  under the 
approximation $h_{1L}^{\perp (1)}(x)=0$.}
\label{F2}  
\end{figure}

\begin{figure}
\begin{minipage}{6cm}
\mbox{\epsfysize=40mm\epsffile{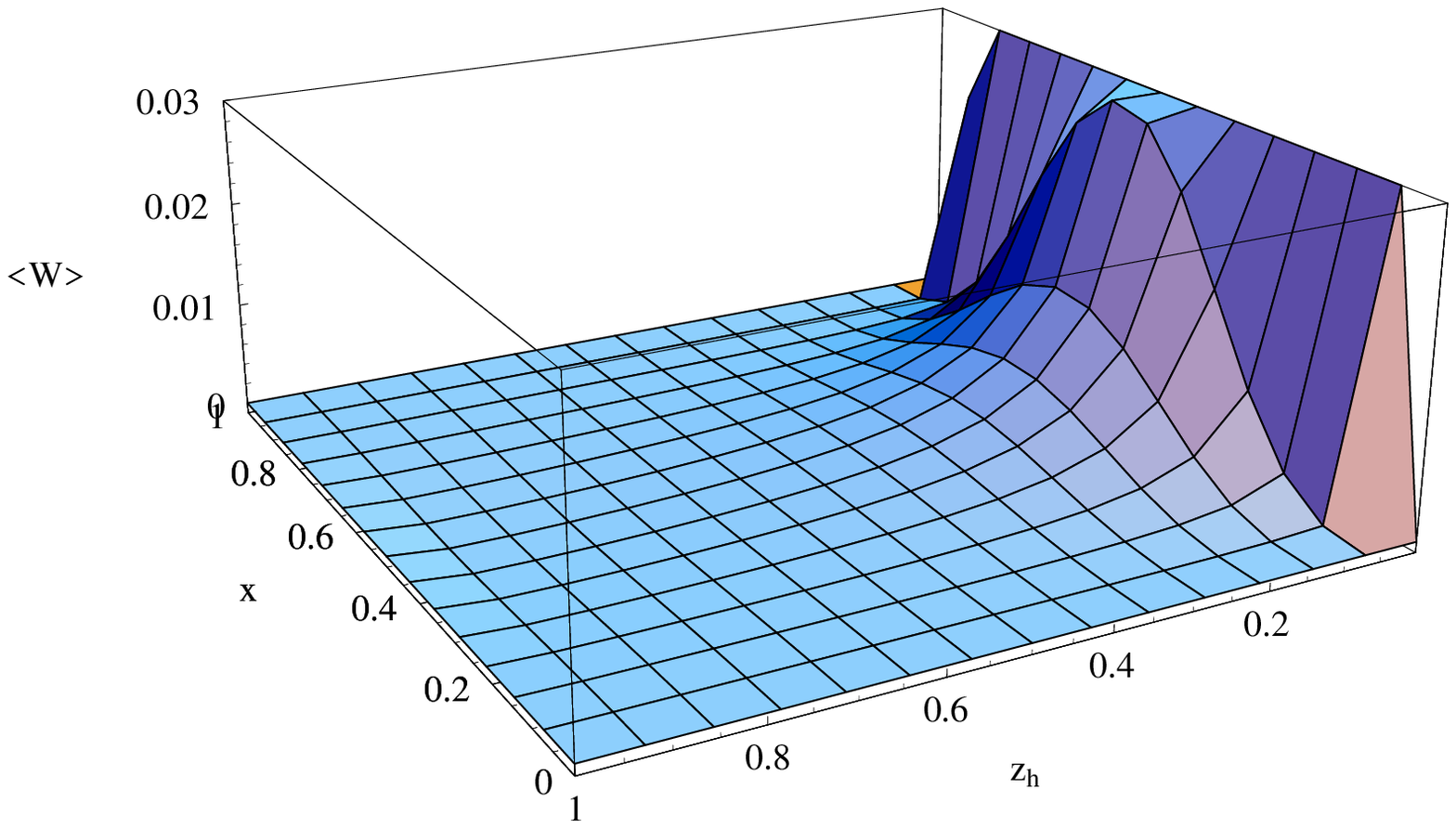}}
\end{minipage}
\begin{minipage}{6cm}
\mbox{\epsfysize=40mm\epsffile{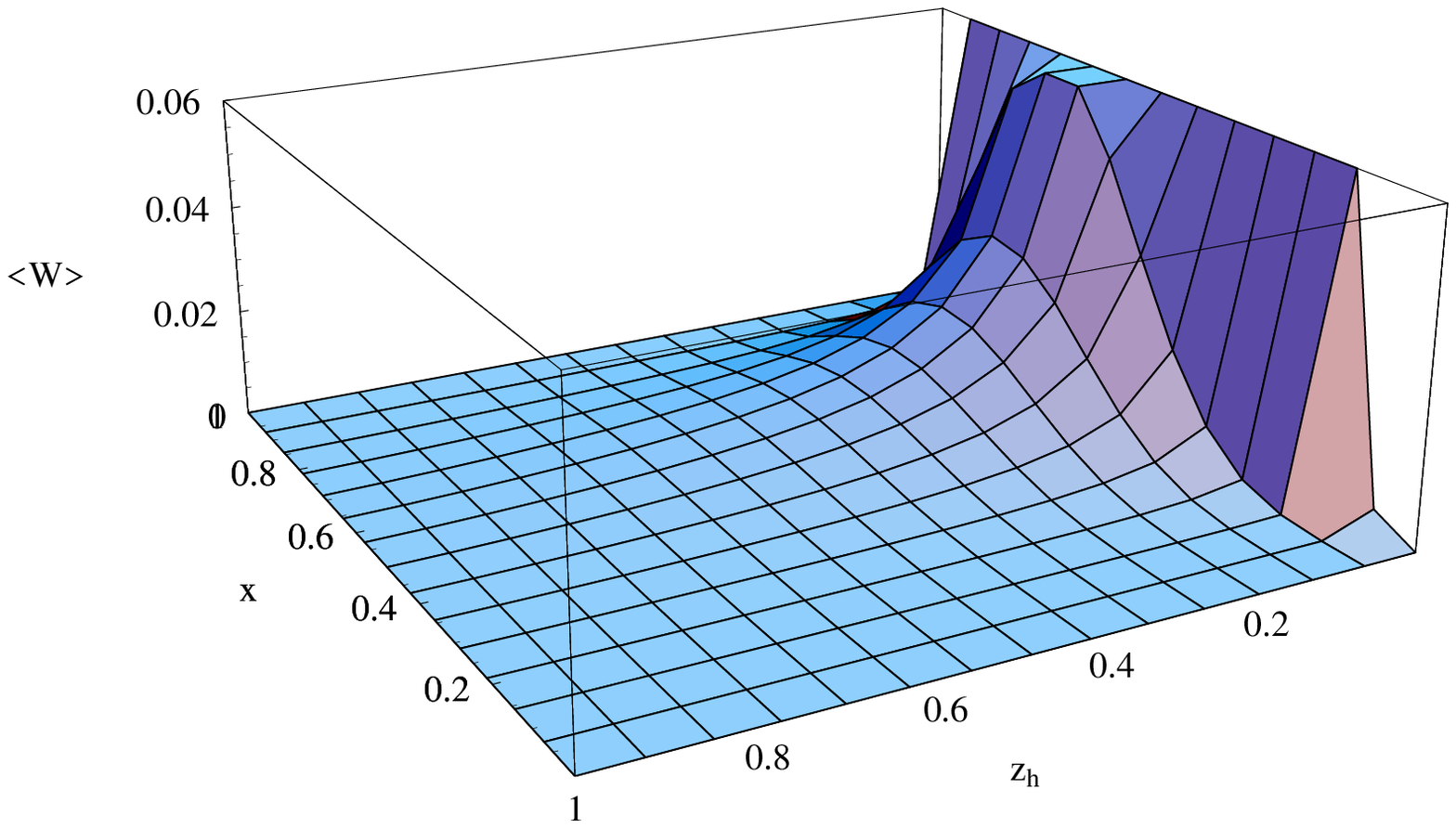}}
\end{minipage}
\caption{ A three-dimensional view of the quantity 
$-\sum _{a,\bar a} e^2_a \Bigl[x_B h_{1L}^{\perp(1)a} (x_B) 
\tilde H ^a (z_h)/z 
- x_B^2  h_L^a (x_B) H_1^{\perp (1) a}(z_h)\Bigr]$, relevant for the 
$\sin (\phi^l _h)$ asymmetry in $\pi^+$ 
production, as obtained 
under the approximation $\tilde h_L = \overline h_L = 0$ (on the left) and 
under the approximation $h_{1L}^{\perp (1)}(x)=0$ (on the right).}
\label{F3}  
\end{figure}

Our results \cite{BM00}, shown in 
Fig.~\ref{F1}, \ref{F2} and \ref{F3} are consistent with the experimental 
data from HERMES as discussed in Ref.~\cite{avakian}.


\newpage

\section*{Conclusions and future perspectives}

Single spin asymmetries are an important tool to learn about 
distribution and fragmentation functions, and ultimately to study the spin 
content of nucleons.
In fact, distribution and fragmentation functions tell us about the internal 
structure of the nucleons and of the role 
their elementary constituents play in accounting for their total spin.
It is then crucial to study those processes in which these functions can be 
exploited. After many years of efforts, both on the experimental and 
theoretical point of view, experimental information on polarized distribution 
and fragmentation functions is now starting 
to come from different sources (HERMES, SMC, SLAC, COMPASS, RHIC and JLAB). 
Thus, some light can be shed, even though we are still far from a completely 
clear picture. This is another little step helping to draw a neater picture 
of the very intriguing ``soft'' physics which governs the hadronic world.

\bigskip
{\small M. Boglione wishes to thank Mirek Finger  
for his invitation to such a nice and interesting conference.
This research project was supported by the Foundation for Fundamental 
Research on Matter (FOM).}

\end{document}